\documentclass[aps, prb, reprint, superscriptaddress, showpacs,longbibliography, preprintnumbers,amsmath,amssymb,twocolumn]{revtex4-2}
\usepackage{amssymb}
\usepackage{amsmath}
\usepackage{bm}
\usepackage[dvips]{graphicx}
\usepackage{dcolumn}
\usepackage{longtable}
\usepackage{color}
\usepackage[version = 4]{mhchem}
\usepackage{comment}

\newcommand{\vect}[1]{\mbox{\boldmath $#1$}}       
\newcommand{\subm}[1]{_{\mathrm{#1}}}  
\newcommand{\spsm}[1]{^{\mathrm{#1}}}  
\newcommand{\CRA}{CeRh$_2$As$_2$~}

\begin{document}
\title{Appearance of $c$-axis magnetic moment in odd-parity antiferromagnetic state in \CRA revealed by $^{75}$As-NMR}

\author{Shiki \surname{Ogata}}
\email{ogata.shiki.86c@st.kyoto-u.ac.jp}
\affiliation{Department of Physics, Kyoto University, Kyoto 606-8502, Japan}
\author{Shunsaku \surname{Kitagawa}}
\affiliation{Department of Physics, Kyoto University, Kyoto 606-8502, Japan}
\author{Katsuki \surname{Kinjo}}
\affiliation{Institute of Multidisciplinary Research for Advanced Materials, Tohoku University, Sendai, Miyagi 980-8577, Japan}
\author{Kenji \surname{Ishida}}
\affiliation{Department of Physics, Kyoto University, Kyoto 606-8502, Japan}

\author{Manuel \surname{Brando}}
\affiliation{Max Planck Institute for Chemical Physics of Solids, D-01187 Dresden, Germany}

\author{Elena \surname{Hassinger}}
\affiliation{Max Planck Institute for Chemical Physics of Solids, D-01187 Dresden, Germany}
\affiliation{Technical University Dresden, Institute for Solid State and Materials Physics, D-01062 Dresden, Germany}

\author{Christoph \surname{Geibel}}
\affiliation{Max Planck Institute for Chemical Physics of Solids, D-01187 Dresden, Germany}
\author{Seunghyun \surname{Khim}}
\affiliation{Max Planck Institute for Chemical Physics of Solids, D-01187 Dresden, Germany}
\date{\today}

\begin{abstract}
\CRA shows the superconducting (SC) multiphase under the $c$-axis magnetic field, which is considered to originate from local inversion symmetry breaking at the Ce site. We reported that the antiferromagnetic (AFM) order is inside the SC phase and that the AFM state disappears at the transition field to the high-field SC phase. However, the magnetic structure in the AFM state has not been clarified yet. In this study, we performed $^{75}$As-NMR measurements in the SC phase in $H\parallel [110]$ to identify the magnetic structure. Comparing the NMR linewidth with $H\parallel c$, we found that the internal magnetic field is oriented to the $c$ axis. This suggests a $\vect{q} = \vect{0}$ $A$-type AFM with the moments parallel to the $c$ axis. We also observed the reduction of the spin susceptibility, which indicates spin-singlet superconductivity in the low-field SC phase. This study provides an important clue to clarify the correlation between the SC multiphase, magnetism, and local inversion symmetry breaking.
\end{abstract}

\pacs{74}
\maketitle
\section{INTRODUCTION}
In conventional superconductors, magnetism and superconductivity are incompatible with each other. In strongly correlated electron systems, on the other hand, superconductivity is often realized near the magnetic ordered phase and the maximum superconducting (SC) transition temperature $T\subm{SC}$ is achieved at the quantum critical point \cite{CeRh2Si2,CeRhIn5,CeCu2(SiGe)2,BaFe2AsP2}. In such superconductors, magnetic fluctuations are considered to be the origin of pair formation \cite{CeRhIn5_T1,BaFe2AsP2_T1,BaFe2AsP2_Pressure_T1}. Therefore, it is extremely important to investigate the relationship between magnetism and superconductivity to identify the mechanism of superconductivity.\par
\CRA is a recently discovered heavy-fermion superconductor with $T\subm{SC}\sim 0.3$ K \cite{hakken}. The crystal structure of \CRA is of tetragonal CaBe$_2$Ge$_2$ type with the space group $P4/nmm$ (No. 129, $D^{7}_{4h}$). There are two crystallographically inequivalent As and Rh sites; As(1) [Rh(1)] is tetrahedrally coordinated by Rh(2) [As(2)], as shown in Fig. 1(a). The $H$-$T$ phase diagram in $H\parallel c$ is quite unusual, showing a clear SC multiphase, while a single SC phase in $H\parallel [110]$. This SC multiphase is quite similar to the theoretically proposed pair-density-wave (PDW) state originating from the locally broken inversion symmetry at the SC layers \cite{T.Yoshida}. This PDW state is intriguing, in which the odd-parity superconductivity with only the spin-singlet SC order parameter is expected to be realized. The PDW state is suggested to exhibit topological superconductivity \cite{Nogaki_topo} and unique responses owing to the SC parity transition \cite{Linear_optical, Szabo_1storder}. \CRA was also reported to exhibit a nonmagnetic phase transition at $T_{0} \sim$ 0.4 K just above $T\subm{SC}$ \cite{hakken, QDW, QDW_theta, QDW_new, QDW_Los}. Although the specific-heat jump at $T\subm{SC}$ is large, the anomaly at $T_{0}$ is rather weak. This transition is considered as an electric quadrupole density-wave (QDW) order of the quasiquartet state consisting of ground and excited doublets coupled by the Kondo effect \cite{QDW, quadrupole_Xray}. In addition to the QDW state, the in-plane antiferromagnetic (AFM) state \cite{T0AFM_theory} and polarized charge distributions of Ce$\spsm{3+}$ ions \cite{miyakeT0} have been discussed as the order parameters below $T_0$. However, there are few experimental reports about order parameters below $T_0$, making it still an open question.\par
In addition to the SC multiphase and QDW phase, AFM order at $T\subm{M} < T\subm{SC}$ was suggested by our nuclear quadrupole resonance (NQR) measurement \cite{Kibune}. This AFM order is realized in the low-field SC phase, and it disappears almost simultaneously with the transition to the high-field SC phase \cite{H||c}. The interplay between the SC multiphase and the AFM state in \CRA has also been studied theoretically \cite{Machida_flop, Wu_symmetryclass, SC-induced_improper_order}. In recent higher-quality single crystals, a first-order transition below $T\subm{SC}$ has also been reported in the specific heat measurement \cite{Poland_AFM_1storder}. On the other hand, recent specific-heat \cite{Poland_AFM_1storder} and muon spin relaxation ($\mu$SR) measurements \cite{muSR} using higher-quality samples suggested that magnetic order emerges from $T_{0}$, indicating that there are still unresolved aspects regarding the magnetic phase. In any case, \CRA is a promising candidate to study the unconventional nonmagnetic, AFM, and SC states as well as their interplays, originating from locally broken inversion symmetry. However, the magnetic structure in the AFM state has not been determined, and the behavior of the SC and AFM phases in $H\parallel [110]$ has not been investigated. Hence, a detailed study of the magnetic state was desired.\par
While the bulk magnetic susceptibility is hindered by the SC diamagnetic shielding effect, nuclear magnetic resonance (NMR) can measure the spin susceptibility in the SC state. In addition, NMR is sensitive to the internal magnetic field at the observed nuclear sites, so the magnetic structure can be discussed by comparing the internal magnetic field at different nuclear sites. In fact, the cancellation of the internal magnetic field at the As(1) site in the NQR measurements \cite{Kibune} leads to two possible magnetic structures: the $A$-type AFM (in-plane ferromagnetic and inter-plane antiferromagnetic) state with magnetic moments parallel to the $c$ axis or the helical state with in-plane moments $\left[ \vect{q} = (\pi , \pi , \pi)\right]$. Therefore, NMR is one of the most powerful techniques to investigate the SC and magnetic properties.\par
In this paper, we report the $^{75}$As-NMR results of \CRA in $H \parallel [110]$. We observed the reduction of the spin susceptibility in the SC state, ensuring our previous report of the formation of spin-singlet pairing in the low-field SC state. This is because the possibility of spin-triplet pairing still remained if the spin susceptibility perpendicular to the $c$ axis was not measured.
The field distribution at the As(2) site is smaller than that in $H\parallel c$ \cite{H||c}, indicating that the internal magnetic field at the As(2) site is parallel to the $c$ axis. From this result, we propose a $\vect{q} = \vect{0}$ $A$-type magnetic structure in the AFM state. This paper provides essential insights into the interplay between magnetism and superconductivity in locally inversion symmetry breaking systems.\par
\section{EXPERIMENTAL}
Single crystals of \CRA were grown using the bismuth flux method \cite{hakken}. Although high quality \CRA single crystals have recently been reported \cite{QDW_new, Poland_growth}, the sample used in this paper is the same early stage sample as in our previous study \cite{Kibune,Kibune_normal,H||c,ogata_normal_ab}. The magnetic field dependence of $T\subm{SC}$ was determined from the second-order derivative of the SC diamagnetic signal with the ac susceptibility measurements using an NMR coil as shown in Fig.1(b). For the NMR measurements, we used a split SC magnet that generates a horizontal field and combined it with a single-axis rotator to apply a magnetic field parallel or perpendicular to the $c$ axis. Low-temperature NMR measurements down to 0.06 K were performed using a $^3$He-$^4$He dilution refrigerator, in which the sample was immersed in the $^3$He–$^4$He mixture to reduce radio-frequency heating during measurements. A conventional spin-echo technique was used for the NMR measurements. The magnetic field was calibrated using $^{63}$Cu (nuclear gyromagnetic ratio $^{63}\gamma _n$/2$\pi$ = 11.289 MHz/T) and $^{65}$Cu ($^{65}\gamma _n$/2$\pi$ = 12.093 MHz/T) NMR signals from the NMR coil. We experimentally confirmed superconductivity immediately after the NMR pulses using a technique reported in previous studies \cite{Heat-up_SRO,UTe2_Knight_shift}. The $^{75}$As-NMR spectra (nuclear spin $I$ = 3/2, $\gamma$/2$\pi$ = 7.29 MHz/T, and natural abundance 100\% ) were obtained as a function of frequency at fixed magnetic fields. Reflecting two crystallographically inequivalent As sites, two $^{75}$As-NMR peaks were observed in all measurement ranges as shown in Figs. 2(a) and 2(b). The site assignment of the NMR peaks was described in a previous paper \cite{Kibune_normal}.\par

   \begin{figure}[htbp!]
   \begin{center}
   \includegraphics[scale=1.05]{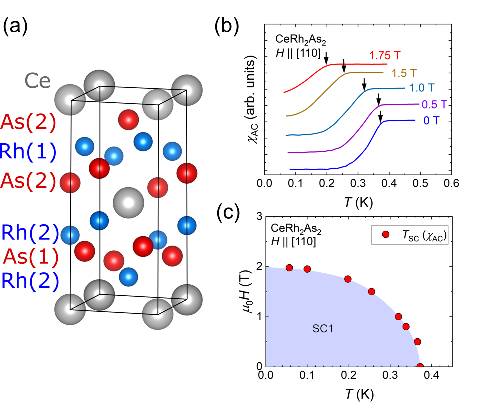}
   \end{center}
   \caption{(a) Crystal structure of CeRh$_2$As$_2$ \cite{VESTA}. (b) Temperature dependence of ac susceptibility measurements in $H\parallel [110]$. The black arrows indicate the superconducting transition temperature $T\subm{SC}$ estimated from the peak of the second-order derivative.(c) $H$-$T$ phase diagram under $H \parallel [110]$.}
   \end{figure}

\section{RESULTS}
To investigate the SC and magnetic properties, particularly the magnetic structure in the AFM state, we performed $^{75}$As-NMR measurements of two As sites. As shown in Figs. 2(a) and 2(b), the NMR frequencies of 0.6 K at $\mu_{0}H \sim 0.8$ T were $f\spsm{As(1)} \sim 11.55$ MHz and $f\spsm{As(2)} \sim 9.56$ MHz, respectively. [See Figs. 4(a) and 4(b) for details of the transition.] The sharp spectrum of $^{65}$Cu-NMR from a NMR coil was also observed. %
The full width at half maximum (FWHM) and the resonance frequency were determined by the fitting, shown with the dashed lines in Figs. 2(a) and 2(b). As for the fitting, we adopted a Gaussian function for the As(1) site and a Lorentzian function for the As(2) site as the fitting function based on the shape of the NMR spectra. Although a theoretical paper predicted the stripe-AFM order at $T_0$ \cite{T0AFM_theory}, we did not find any change both in the Knight shift and FWHM within the resolution at $T_0$. At low temperatures, the spectrum of the As(1) site became asymmetric. Thus, we used an asymmetric Gaussian function with an asymmetrization factor $\alpha$: $\frac{1}{\sqrt{2\pi}w}\mathrm{exp}\left[-\frac{(x-\mu)^2}{2[w + \alpha(x-\mu)]^2}\right]$\cite{asymmetric_gauss1}. Such asymmetric broadening may be caused by an internal field at the As(1) site, for example, field-induced ferromagnetic components with the AFM state. Since the hyperfine coupling constant of the As(1) site is negative in $H\parallel [110]$, ferromagnetic components broaden the NMR spectra in the low-frequency direction. We adopted the peak frequency as the Knight shift, and estimated the Knight shift by numerical calculations with a diagonalized nuclear Hamiltonian.
Below $T\subm{SC}$, the NMR spectra broaden due to SC diamagnetism. However, a significant site-dependent broadening, observed in $H\parallel c$ below 0.25 K \cite{H||c}, was not observed in $H\parallel [110]$. The Knight shifts for the As(1) and As(2) sites are denoted by $K\spsm{As(1)}$ and $K\spsm{As(2)}$ in Fig. 2(d), respectively. In the normal state, the Knight shifts of both sites are constant below 1 K. Below $T\subm{SC}$, $K\spsm{As(2)}$ decreased, but $K\spsm{As(1)}$ slightly increased. Note that the hyperfine coupling constant $A_{\mathrm{hf},[110]}$ of the As(1) site is negative while that of the As(2) site is positive \cite{ogata_normal_ab}, thus we can say that we observed a decrease of the spin susceptibility at both sites in the SC state. Considering that the spin susceptibility also decreases in $H\parallel c$ \cite{H||c}, this indicates that \CRA is in the spin singlet pairing in the low-field SC state.

   \begin{figure}[htbp!]
   \begin{center}
   \includegraphics[scale=0.98]{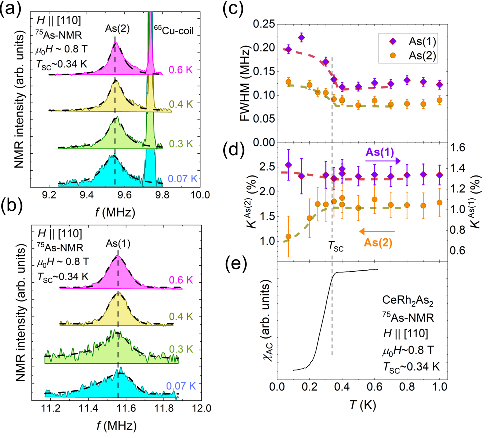}
   \end{center}
   \caption{NMR measurements in CeRh$_2$As$_2$. The temperature evolution of the NMR spectra of (a) the As(1) site and (b) the As(2) site at 0.8 T for $H\parallel [110]$. The dashed vertical lines indicate the NMR frequency of the normal-state signal. The dashed curves indicate the results of a fitting. Temperature dependence of (c) full width at half maximum (FWHM), (d) Knight shift at the As(1) site $K\spsm{As(1)}$, and As(2) site $K\spsm{As(2)}$ determined by the NMR spectrum at 0.8 T. $T\subm{c}$ is indicated by the black dashed line. The colored dashed lines in (c) and (d) are guides to the eye. (e) Temeperature dependence of ac magnetic susceptibility at 0.8 T for $H\parallel [110]$.}
   \end{figure}

From the obtained NMR spectra, we estimated the magnetic field distribution $\Delta\mu_{0}H$ at each As site arising from the staggered internal field in the AFM state. In $H\parallel [110]$, the effect of the nuclear quadrupole interaction cannot be ignored unlike the case of the central peak (1/2 $\leftrightarrow$ $-$1/2 transition) in $H\parallel c$. Therefore, we used the slope of the resonance frequency against the magnetic field obtained by numerical calculations with a diagonalized nuclear Hamiltonian as shown in Fig. 3(a). From the slope, the effective gyromagnetic ratio at each site is estimated as 13.47 MHz/T for the As(1) site and 3.33 MHz/T for the As(2) site in the measurement field region, respectively. Using this coefficient, the increase of FWHM in the SC state is converted to the increase of the magnetic field distribution $\Delta\mu_{0}H$, as shown in Fig. 3(b). The result of the central peak in $H\parallel c$ in the previous study \cite{H||c} is also shown in Fig. 3(c).\par

   \begin{figure*}[htbp!]
   \begin{center}
   \includegraphics[scale=0.9]{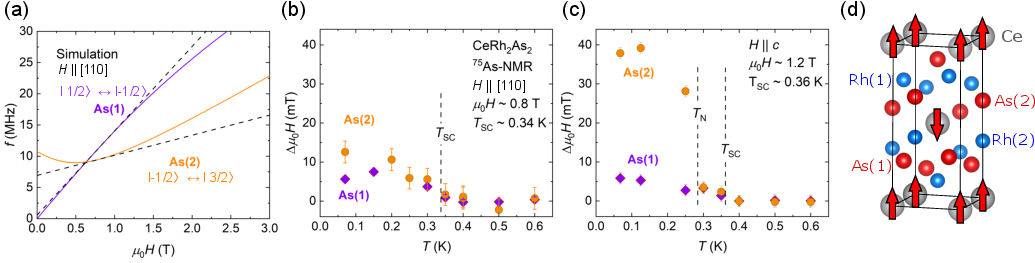}
   \end{center}
   \caption{(a) Magnetic field dependence of NMR frequency of each As site obtained from NMR spectrum simulation. The dashed lines indicate the result of linear fitting at 0.8 T. (b) The temperature dependence of the field distribution at As sites at 0.8 T for $H\parallel [110]$ estimated from simulation. (c) The temperature dependence of the field distribution at As sites at 1.2 T for $H\parallel c$ (estimated from Ref. \cite{H||c}). (d) The possible magnetic structure with $H\subm{int} \parallel c$ at the As(2) site. }
   \end{figure*}
\section{DISCUSSION}
Comparing the result in $H\parallel [110]$ and $H\parallel c$, we determine the orientation of the internal field in the AFM state. The staggered internal field in the AFM state causes the splitting of the NMR spectra in principle, and makes the spectra broaden in observation. Here, the broadening (or splitting) by the internal field depends on the orientation relative to the applied field; the broadening is largest when the internal field is parallel to the applied field. As is clear from Figs. 3(b) and 3(c), the field distribution $\Delta\mu_{0}H$ at the As(2) site is larger in $H\parallel c$, while that at the As(1) site is almost the same in both field directions. This indicates that the internal magnetic field at the As(2) site is oriented to the $c$ axis and the magnitude is around 16 mT, which is estimated from half of the difference of the field distribution between the two As sites in $H\parallel c$. The magnetic structures, in which the internal magnetic field cancels at the As(1) site, are pinned down to the two magnetic structures, as discussed in Ref. \cite{Kibune}. In these two structures, the one shown in Fig. 3(d) is the most plausible structure, considering the orientation of the internal magnetic field at the As(2) site. Thus, we propose that the magnetic structure is the $\vect{q} = \vect{0}$ $A$-type AFM state with magnetic moments $\mu \parallel c$. Using the hyperfine coupling constant of the As(2) site in $H\parallel c$ $( \sim$ 0.27 T/$\mu\subm{B})$ \cite{Kibune_normal, ogata_normal_ab}, the ordered moment is estimated to be on the order of 0.1 $\mu\subm{B}$. Note that the field distribution at the As(2) site in the zero-field NQR at the lowest temperature (0.1 K) \cite{Kibune} is larger than that in the $c$-axis 1.2 T NMR (0.07 K) \cite{H||c}, suggesting that the magnetic moment is reduced by the $c$-axis field. Such a small ordered moment is consistent with the absence of the static magnetic scattering in elastic neutron diffraction measurements \cite{AFM_fluctuation_APSmeet}. Here, in systems with small magnetic anisotropy, the direction of the staggered moments is expected to reorient perpendicular to the $c$ axis under the $c$-axis field. In such cases, however, the increase of the field distribution of 16 mT requires a ten times larger ordered moment ($\sim 1 \mu\subm{B}$), which is inconsistent with the low-temperature linewidth of the NQR spectra \cite{Kibune}. It is noted that this magnetic structure seems to be incompatible with the anisotropy of the spin susceptibility in the normal state and the presence of the two-dimensional (2D) $XY$-type AFM fluctuation \cite{Kibune_normal,AFM_fluctuation_APSmeet}. It is considered that the orientation of the ordered moments is determined by the spin-orbit coupling or the interaction between magnetic moments. Since this study was performed on an early stage sample, it is important to measure the sample dependence using the recently reported higher-quality samples \cite{QDW_new, Poland_growth}.
\par
Certain AFM states in the locally inversion symmetry breaking crystals break global inversion symmetry, which are called odd-parity multipolar states \cite{multipole_BaMn2As2, multipole_zigzag_SC, monopole_Nicola, cross-correlation_monopole}. In fact, the $\vect{q} = \vect{0}$ $A$-type AFM state with $\mu \parallel c$ is equivalent to the case of the magnetic monopole order \cite{multipole_zigzag_SC, monopole_Nicola}. Cross-correlated responses such as the electromagnetic effect are expected in the magnetic monopole order \cite{cross-correlation_monopole}. Superconductivity coexisting with odd-parity multipoles is very rare. Therefore, \CRA provides a promising platform to study the relationship between superconductivity and unconventional multipoles.
\par
Next, we discuss the change in the Knight shift below $T\subm{SC}$. In a conventional spin-singlet superconductor, the spin susceptibility decreases and becomes almost zero at $T \rightarrow 0$ K due to the disappearance of the electron spin degrees of freedom as a pair formation. The spin susceptibility can be obtained from the Knight-shift measurement, but the Knight shift also includes a temperature-independent orbital component associated with Van-Vleck susceptibility, and the SC diamagnetic effects. Therefore, the temperature dependence of the observed Knight shift $\Delta K$ can be divided into the following three contributions,
\begin{flalign}
\Delta K = K\subm{normal} + \delta K\subm{spin} + K\subm{dia},
\end{flalign}
where $K\subm{normal}$ is the temperature dependent Knight shift in the normal state, $\delta K\subm{spin}$ is the change in the spin component of the Knight shift ascribed to the pair formation, and  $K\subm{dia}$ is the SC diamagnetic component. In the SC state, the Knight shift decreases due to the SC diamagnetic shielding effect, and the value at 0 K is approximately expressed as \cite{Kdia}
\begin{flalign}
K\subm{dia} = -\frac{H\subm{c1}}{H}\frac{\ln{\left(\frac{\beta \lambda_d}{\sqrt{e}\xi}\right)}}{\ln{\kappa}}.
\end{flalign}
Here, $H\subm{c1}$ is the lower critical field, $\xi$ is the Ginzburg-Landau (GL) coherence length, $\beta$ is a factor that depends on the vortex structure and is 0.38 for the triangular vortex lattice, $\lambda_d$ is the distance between the vortices and is calculated using the relation $\phi_0 = \frac{\sqrt{3}}{2}\lambda^{2}_{d}(\mu_{0}H\subm{ext})$, $e$ is Euler's number, and $\kappa$ is the GL parameter. From the orbital limiting field $\mu_{0}H\subm{orb} \sim 8$ T, and the thermodynamic critical field $\mu_{0}H\subm{c} = 31$ mT \cite{hakken}, $\mu_{0}H\subm{c1} = 0.625$ mT , $\xi = 6.42$ nm, and $\kappa = 182$ were adopted. At the measured magnetic field of $\mu_{0}H \sim 0.8$ T, $K\subm{dia}$ is estimated to be around 0.01\%. Despite this value being several times smaller than the error bar of the measurement and does not have a significant impact in the study, this value is subtracted from here on.\par

\begin{figure*}[htbp!]
   \begin{center}
   \includegraphics[scale=1.05]{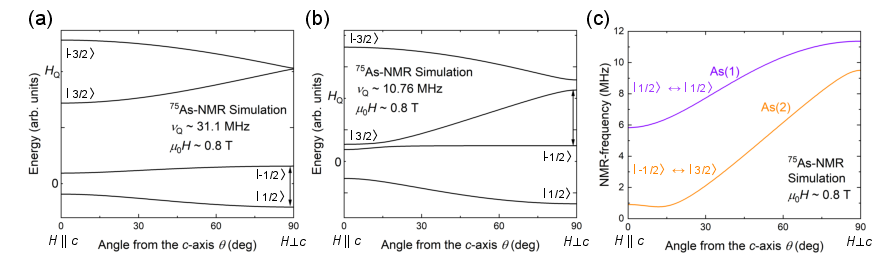}
   \end{center}
   \caption{The polar angle $\theta$ dependence of the energy level of the $^{75}$As nuclear spin simulated by diagonalizing nuclear Hamiltonian for (a) the As(1) site and (b) the As(2) site. $H\subm{Q} = h\nu\subm{Q}$ is the energy of the nuclear quadrupole interaction ($H = 0$). The arrows represent the transitions observed in the experiment. (c) $\theta$ dependence of the NMR frequency at each As site.}
\end{figure*}

To quantitatively discuss the reduction of the Knight shifts in Fig. 2(d), we compared $H\subm{c2}$ with the Pauli-limiting field $H\subm{P}$ estimated from the change in the Knight shift as follows. In a heavy-fermion spin-singlet superconductor, $H\subm{c2}$ in the low temperature region is determined by the magnetic field where the energy gain by Zeeman-splitting effect is as high as the SC condensation energy: It is called the Pauli depairing effect. It is well known that the Pauli limiting field $H\subm{P}$ for spin-singlet superconductors is expressed by the following simple equation with the decrease in the spin susceptibility $\delta \chi$ ascribed to singlet-pair formation:
\begin{flalign}
\frac{1}{2}\delta \chi\; \mu_{0}H\subm{P}(0)^{2} = \frac{1}{2}\mu_{0}H^{2}\subm{c}.
\end{flalign}
This equation yields $\mu_{0}H\subm{P}(0) = \mu_{0}H\subm{c}/\sqrt{|\delta \chi |}$, where $\delta \chi$ can be estimated from $\delta K\subm{spin}$ as $\delta \chi = A\subm{hf} \delta K\subm{spin}$. When we assume that superconductivity in $H\parallel [110]$ is destroyed by the Pauli depairing effect, we evaluate the corresponding Knight-shift reduction to be $\delta K\spsm{As(1)}\subm{spin} \sim -0.04$\% (slightly increase) and $\delta K\spsm{As(2)}\subm{spin} \sim 0.03$\%, respectively. Thus, for the As(1) site, $\delta K\subm{spin}$ is the same order as the error bar, but for the As(2) site, experimental $\delta K\subm{spin} \sim$ 0.4\% is one order larger than the estimation. Therefore, the resonance frequency of the As(2) site might decrease by an effect other than the decrease associated with the SC spin susceptibility.\par
Here, we consider the origin of the large reduction of $K\spsm{As(2)}$. In the above discussion, we assumed that the change in NMR frequency is totally due to that in the Knight shift, but change in NQR frequency $\nu\subm{Q}$ and the internal magnetic field also shift the resonance frequency. However, it was suggested that the temperature evolution of $\nu\subm{Q}$ was negligibly small \cite{Kibune}, and thus we consider the effect of the internal magnetic field in the AFM state. As shown in Fig. 3(d), the internal field at the As(1) site cancels out, which is in good agreement with the fact that a large $\Delta K$ is observed only at the As(2) site. The magnitude of the effective field at the nucleus, $H\subm{eff} = \sqrt{H\subm{int}^{2} + H\subm{ext}^2}$, is almost the same as that of the external field $H\subm{ext}$ when the internal field $H\subm{int}$ is small. However, the direction of $H\subm{eff}$ is slightly tilted to the $c$ axis due to the appearance of the internal field along the $c$ axis, which can change the resonance frequency of the As(2) site. Figure 4 shows the polar angle $\theta$ dependence of the nuclear level and NMR frequency simulated by diagonalizing the nuclear Hamiltonian. As can be seen in Fig.4(c), the NMR frequency at the As(2) site decreases as $H\subm{eff}$ shifts from $H\perp c$. When we adopt that the internal magnetic field at the As(2) site is 16 mT along the $c$ axis, the resonance frequency decreases by 11.5 kHz, corresponding to $\Delta K \sim 0.43$\%. Thus, it is noted that the effect of the internal magnetic field is involved in the overestimation of the Knight shift. \par
\section{CONCLUSION}
We performed $^{75}$As-NMR measurements on \CRA in order to investigate the in-plane spin susceptibility in the low-field SC state and the direction of the internal magnetic field in the AFM state coexisting with superconductivity. By combining with previous results, the spin susceptibility in all directions decreases in the SC state, indicating a spin-singlet state. The excess decrease in the Knight shift at the As(2) site in $H\parallel [110]$ could be interpreted as the effect induced by the internal magnetic field along the $c$ axis. In addition, the field distribution due to the internal magnetic field at the As(2) site is much smaller than that in $H\parallel c$, suggesting that the internal magnetic field is oriented to the $c$ axis. We point out the possible magnetic structure shown in Fig. 3(d) by considering it together with the cancellation of the internal magnetic field at the As(1) site. Quite recently, $\mu$SR measurements detected a static field below $T_{0}$, which has comparable in-plane and $c$-axis components with a weakly anisotropic magnetic structure \cite{muSR}. This seems to be different from our NMR structure, but the sample quality and experimental conditions are different between the NMR and $\mu$SR measurements. Further experiments are needed to settle the issue. In any cases, our results should be fundamental for clarifying the relationship between the field-induced SC multiphase and the magnetism in CeRh$_2$As$_2$.\par

\vskip.5\baselineskip
\begin{acknowledgments}
This work was partially supported by the Kyoto University LTM Center and Grants-in-Aid for Scientific Research (KAKENHI) (Grants No. JP19K14657, No. JP19H04696, No. JP20KK0061, No. JP20H00130, No. JP21K18600, No. JP22H04933, No. JP22H01168, and No. JP23H01124, No. JP23K19022, and No. JP24KJ1360). This work was also supported by JST SPRING (Grant No. JPMJSP2110) and research support funding from the Kyoto University Foundation, and ISHIZUE 2024 of Kyoto University Research Department Program, and Murata Science and Education Foundation. C. G. and E. H. acknowledge support from the DFG program Fermi-NESt through Grant No. GE 602/4-1. Additionally, E. H. acknowledges funding by the DFG through CRC1143 (Project No. 247310070) and the Würzburg-Dresden Cluster of Excellence on Complexity and Topology in Quantum Matter—ct.qmat (EXC 2147, Project ID 390858490).
\end{acknowledgments}

\end{document}